# Precision Measurement of the Electron/Muon Gyromagnetic Factors

A. M. Awobode

Department of Physics, University of Illinois, Urbana-Champaign, IL 61801, U.S.A

**Abstract**
Clear, persuasive arguments are brought forward to motivate the need for highly precise measurements of the electron/muon orbital g, i.e. $g_L$. First, we briefly review results obtained using an extended Dirac equation, which conclusively showed that, as a consequence of quantum relativistic corrections arising from the time-dependence of the rest-energy, the electron gyromagnetic factors are corrected. It is next demonstrated, using the data of Kusch & Foley on the measurement of $(\delta_S - 2\delta_L)$ together with the modern precise measurements of the electron $\delta_S$ ($\delta_S \equiv g_S - 2$), that $\delta_L$ may be a small ($-0.6 \times 10^{-4}$), non-zero quantity, where we have assumed Russel-Saunders (LS) coupling and proposed, along with Kusch and Foley, that $g_S = 2 + \delta_S$ and $g_L = 1 + \delta_L$. Therefore, there is probable evidence from experimental data that $g_L$ is not exactly equal to 1; the expectation that quantum effects will significantly modify the classical value of the orbital g is therefore reasonable. Finally we show that if, as suggested by the results obtained from the modified Dirac theory, $\delta_S$ and $\delta_L$ depend linearly on a dimensionless parameter $\Delta$ such that the gyromagnetic factors are considered corrected as follows; $g_S = 2(1 + \Delta)$ & $g_L = 1 - \Delta$, then the Kusch-Foley data implies that the correction $\Delta \approx 1.0 \times 10^{-3}$; it is noteworthy that $\Delta$ is of the same order of magnitude as the measured $(g_S - 2)/2$ which, to five places of decimal, is equal to $1.12 \times 10^{-3}$. Thus, available spectroscopic data indicate that $g_S$ and $g_L$ may both be significantly modified, such that $g_S$ is increased by $2\Delta$, while $g_L$ is decreased by $\Delta$, the quantity $2\Delta$ being equal to the precisely measured $g_S - 2$. Modern, high precision measurements of the electron and muon orbital $g_L$ are therefore required, in order to properly determine by experiments the true value of $g_L - 1$, perhaps to about one part in a trillion as was recently done for $g_S - 2$.

PACS 12.20.-m, 12.20.Fv

## 0. Introduction.

In search of new physics beyond the standard model, QED has been stringently tested to extremes of precision especially in the measurement of the Lamb shift and the anomalous magnetic moment of spin-1/2 particles. Measurements have been carried out on single "isolated" electrons [1,2,3], atomic (bound) electrons [4,5], "free" (unbound) electrons [6,7], and relativistic muons [8,9]. So far QED has brilliantly survived all the tests, except for some deviations from experiments in the value of the intrinsic magnetic moment of muons [10,11], which may perhaps be attributed to the influence of interactions other than the electromagnetic. Nevertheless, there is hope that further tests will eventually be fruitful, and that new physics will become revealed.

Departing from the traditional, high-precision measurement of the Lamb shift or the intrinsic g factor, it is hereby proposed that a qualitatively different kind of measurement (i.e., of the orbital g), be considered. There are reasons, both experimental and theoretical, to suggest that the orbital g, when measured to very high levels of precision, will be found not to be exactly equal to 1. Thus, the new question to be addressed by experiment may succinctly be put as follows; if we assume that $g_L = 1 + \delta_L$, is $|\delta_L| > 0$ ?

A preliminary answer to this question comes from the classic measurement by Kusch and Foley [12], of the quantity $(\delta s - 2\delta_L) \neq 0$ considered together with the recent measurements of $\delta s$ [1,2,3]. It will be shown in what follows that using experimental evidence presently available, it cannot be asserted with certainty that (on empirical grounds) $\delta_L = 0$. It is indeed highly probable that $\delta_L \neq 0$; that is, a measurable, small but finite quantity $\delta_L$ exists, such that the orbital magnetic moment of the electron differs from the classical value $g_L = 1$.

Quite apart from the above-mentioned empirical evidence, there are also theoretical reasons to expect that the orbital gyro-magnetic factor $g_L$ when precisely measured will differ from exact unity. The electron, unlike the proton or neutron, is a point particle which, however, interacts with its own radiation field as described by the wonderfully successful QED. One of the results of this self-interaction and its interaction with the QED vacuum is a correction of the electron charge and mass which had hitherto appeared in electron theories as measurable, fixed parameters. A consequence of these interactions is the correction of the spin magnetic moment of the electron by a factor which has been astonishingly confirmed by subsequent precision measurements [1-7,12]. By considering the consequences of a time-dependent correction to the rest-energy in the Dirac electron theory, it has also been shown that $g_S$ and $g_L$ are both subject to significant modifications; hence a non-zero $\delta_L$ does not necessarily imply that the electron has a substructure.

Here we shall briefly review first the corrections ($\delta s$ and $\delta_L$) to the gyromagnetic factors, produced as a result of the interaction between an electron and a weak magnetic field, as described by an extended Dirac equation, and then follow the discussion with the result obtainable when the classic Kusch-Foley experimental measurement of $(\delta s - 2\delta_L)$ is combined with the precise measurements of $\delta s$, which has since become available. Finally, conclusions are drawn regarding the significance of the various experimental and theoretical observations[1], and a case is made for new, high precision, direct measurements of $\delta_L$ and/or $(\delta s - 2\delta_L)$ using modern experimental techniques, procedures and apparatus.

## 1. Quantum Corrections & the Gyro-magnetic Factors

The rest-energy of a relativistic electron is neither a conserved quantity nor a constant of the motion. From these and some other fundamental physical considerations, it has been shown that the rest-energy has a time-dependence [13]

$$m(t) = m - m\exp(2i\alpha \cdot pt/\hbar) \quad ; \quad c = 1 \tag{1.1}$$

---

[1] An **Appendix** is included in which a heuristic argument is used to obtain expressions relating corrections to the g-factors $\delta_S$ and $\delta_L$, and a formula for $\delta_J$ is derived.

where m is the rest mass, **α** is the Dirac matrix-vector, **p** is the momentum and t is the coordinate time

Let it be accepted as a testable hypothesis, that the known deficiencies of the Dirac theory may be corrected by substituting for the constant rest-energy term, the time-dependent rest-energy function given above. The resulting time-dependent equation [14] is,

$$i\hbar\frac{\partial\psi}{\partial t} = [c\alpha \cdot p + \beta mc^2 - \beta mc^2 \exp(2i\alpha \cdot pct/\hbar)]\psi \qquad (1.2)$$

which allows a consistent formal interpretation of the unusual properties of the Dirac relativistic electron theory.

Now, if the electron interacts with a slowly varying magnetic field B, the Hamiltonian in the above equation, when written in a Schroedinger-type, time-independent form [15], is

$$H = c\alpha \cdot \pi + \beta mc^2 - \frac{i\beta mc(\hbar\varepsilon)(\alpha \cdot \pi)}{4|p|^2} \qquad (1.3)$$

where the exponential term in (1.2) has been re-expressed in the limit $t \to \infty$. By taking the non-relativistic limit of the eigenvalue equation of the above Hamiltonian, an extended version of the Pauli equation is obtained, which may be expressed as [16],

$$E_A \psi_A = \left[\frac{\pi^2}{2m} - \frac{\pi^2 \hbar^2 m^2 \varepsilon^2}{2m\, 4^2 p_A^4} - \frac{e\hbar}{2mc}\left(1 + \frac{\hbar^2 m^2 \varepsilon^2}{4^2 p_A^4}\right)\sigma \cdot B\right]\psi_A \qquad (1.4)$$

where in the term $(\hbar^2 m^2 \varepsilon^2 / 4^2 p_A^4)$, $\varepsilon$ is the frequency of "cyclical" photon associated with the Dirac particle [14] and $p_A$ is the momentum of the particle. The second and fourth terms are new; they have arisen as a consequence of the time-dependent rest-energy term in eqn(1.2). We can interpret the terms by considering the energy of a magnetic moment of a dipole in a weak, static magnetic field B, i.e.,

$$H_{m,S} = -\mu_S \cdot B = \frac{eg_S}{2mc} S \cdot B \quad ; \quad H_{m,L} = \mu_L \cdot B = \frac{eg_L}{2mc} L \cdot B \qquad (1.5)$$

from which we deduce that the electron gyromagnetic factors have new values given as follows;

$$g_S = 2(1+\Delta) \quad : \quad g_L = 1 - \Delta \qquad (1.6)$$

where $\Delta$ is the correction to the g-factors due to the addition to the rest energy term in the Dirac equation. The quantity $\Delta$ is the numerical value of the term $(\hbar^2 m^2 \varepsilon^2 / 4^2 p^4)$ contained in eqn(1.4). On evaluation, it is found in the relativistic regime that $\Delta \approx 1.6 \times 10^{-3}$ which, to four places of decimal, is nearly equal to the measured anomalous part of the intrinsic g-factor, i.e., $a_e = 1.2 \times 10^{-3}$.

If we adopt the suggestion by Kusch & Foley, sequel to their measurements, that the gyromagnetic factors $g_S$ and $g_L$ may be corrected as follows [12];

$$g_S = 2 + \delta_S \quad ; \quad g_L = 1 + \delta_L \qquad (1.7)$$

then from (1.6) above, $\delta_S = 2\Delta$ and $\delta_L = -\Delta$. It is necessary however, to note that there are other possible forms which the corrections to the g-factors can take. For example, it is not impossible that $\delta_S \neq 0$ while $\delta_L = 0$ and vice-versa. Nevertheless, all the various possibilities can be expressed as

$$g_S = 2(1+\Delta) \quad ; \quad g_L = 1 - \lambda\Delta \qquad (1.8)$$

where $\lambda$ is a real number, having taken into considerations the fact that it has been experimentally and theoretically shown that $g_S = 2(1+\Delta)$ where, to four places of decimal, $\Delta = 1.2 \times 10^{-3}$. Thus the choice $\lambda = 1$ refers to the expressions in eqn (1.6) while $\lambda = 0$ implies that $g_L = 1$; the number $\lambda$ can take any real values other than 0 or 1.

We shall, in the following sections, ascertain the compatibility of the above suggestions with available spectroscopic data and draw significant conclusions regarding the true (or probable) value of the orbital g.

## 2. Atomic Spectroscopy Experiments on Bound Electrons

We shall now consider in greater detail, the classic Kusch-Foley experiment [12] in which the quantity ($\delta_S - 2\delta_L$) is determined from the study of the Zeeman splitting of two different atomic states which allowed the calculation of the ratio of the $g_J$ factors corresponding to the states. It will be shown that in the light of modern, highly-precise measurements of $\delta_S$, the Kusch-Foley data suggests that $\delta_L$ may not be exactly zero.

Using the atomic beam magnetic resonance technique, frequencies associated with the Zeeman splitting of the $^2P_{1/2}$ and $^2P_{3/2}$ states of Ga were measured by Kusch & Foley, in order to determine the $g_J$ values corresponding to the states. Since each of these states in Ga may be separately subject to configuration interaction perturbations, Kusch and Foley suggested that the interpretation of the result may be rendered unclear, thus making necessary a new determination of the ratio of the $g_J$ value of Na in the $^2S_{1/2}$ state to that of Ga in the $^2P_{1/2}$ state. From any experiment in which the ratio of the $g_J$ values of the states is measured, it is possible to determine only the quantity ($\delta_S - 2\delta_L$) if LS coupling is assumed[2]. In order to determine either $\delta_S$ or $\delta_L$, it was therefore necessary for Kusch & Foley to make certain assumptions concerning the nature of $\delta_S$ and $\delta_L$, some of which had no obvious experimental basis or support.

A determination of the ratio of $g_J$ from the $^2P_{3/2}$ and the $^2P_{1/2}$ states of Ga gave a value $g_{3/2}/g_{1/2} = 2.00344 \pm 0.00012 = 2 + A$, different from the expected value of 2. The discrepancy, if the states are assumed correctly described by Russel-Saunders (LS) coupling, can be attributed to a change from their accepted values, in the electron intrinsic moment and/or the orbital moment. Kusch and Foley reasoned that if the intrinsic g is changed to $g_S = 2 + \delta s$ and the orbital g becomes $g_L = 1 + \delta_L$, then $A =$ (3/2)$\delta s$ − 3$\delta_L$ or $\delta s - 2\delta_L = 0.00229 \pm 0.00008$. Since this measurement, as earlier remarked, did not permit an independent evaluation of $\delta s$ and $\delta_L$, an attempt was made to account for the discrepancy between the expected and measured values of $g_S$ and $g_L$ by either, (a) setting $\delta_L = 0$ and hence having $\delta_S = 0.00229 \pm 0.00008$, such that $g_s = 2.00229 \pm 0.00008$ & $g_L = 1$, or (b) setting $\delta_S = 0$ and thus putting $\delta_L = 0.00114 \pm 0.00004$, such that $g_S = 2$ $and$ $g_L = 0.99886 \pm 0.00004$.

In a subsequent version of the experiment, the determination of the ratio of the $g_J$ values of Na $^2S_{1/2}$ and of Ga $^2P_{1/2}$ states gave the value $Na\, g_{1/2}/Ga\, g_{1/2} = 3.00732 \pm 0.00018$, which differed significantly from the expected value of 3, thereby making $\delta_S - 2\delta_L = 0.00244 \pm 0.00006$. In order to find $\delta_S$, it was necessary for Kusch & Foley to reason that "if on the basis of the correspondence principle, we set $\delta_L$ equal to zero", then from the first experimental result, $g_S = 2.00229 \pm 0.00008$ and from the second experiment, $g_S = 2.00244 \pm 0.00006$

When compared, these experimentally measured values of $g_S$ are roughly in good agreement with the value theoretically calculated (to first order in the structure constant α) using QED, i.e., $g_S = 2(1 + \alpha/2\pi)_e^{QED} = 2.00232$. Let us, however, reconsider the Kusch-Foley measurements in the light of modern experiments. Using different techniques, more precise values of $\delta_S$ have since been measured, which focused exclusively on the spin g-factor $g_S$; thus we now have independent values of $\delta_S$ and can therefore determine $\delta_L$ within the limits of the accuracy of the earlier experiments, without arbitrarily setting $\delta_L$ equal to zero. Recent measurements give the anomalous part of the intrinsic gyro-magnetic factor as [3,19],

$$a_e \equiv \left((g_S - 2)/2\right)_e = (1159\,652.4 \pm 0.2) \times 10^{-9} \quad (2.1)$$

which implies that, to five places of decimal, $\delta_S = 2a_e = 0.00232$ exactly. Thus by substituting this value for $\delta s$ in the measured ($\delta_S - 2\delta_L$) we can evaluate the value of $\delta_L$.

---

[2] It appears reasonable to assume LS coupling for the states of Na and Ga considered, following a standard criterion [18], which requires that for LS coupling the intervals in the fine structure must be small compared with the differences between levels with different L and S.

We shall consider two special cases in which (a) the corrections $\delta_S$ and $\delta_L$ are assumed independent of one another, and (b) the quantities $\delta_S$ and $\delta_L$ are assumed dependent, and related via a parameter.

**(a). Independent Correction Factors $\delta_S$ & $\delta_L$:** Assuming in consonance with Kusch & Foley, that their experiments measured $(\delta_S - 2\delta_L)$ for the electron considered essentially free from bound state effects, then substituting $\delta_S = 0.00232$ into this expression we find that $\delta_L = 0.00002 \pm 0.00004$ or $\delta_L = (0.2 \pm 0.4) \times 10^{-4}$, from the first measurement. Thus it is observed that $\delta_L$ could lie between $-0.2 \times 10^{-4}$ and $+0.6 \times 10^{-4}$. Noting the relative magnitude of the quantity and that of its error, it does not appear feasible from this result, to assert with certainty that $\delta_L = 0$.

Likewise, substituting $\delta_S = 0.00232$ into the second measurement gives $\delta_L = -0.00006 \pm 0.00003$ which may similarly be re-expressed as $\delta_L = (-0.6 \pm 0.3) \times 10^{-4}$. Here we observe that, though close to the edge of precision, the data shows that $\delta_L$ is a probable, small but finite quantity which, with modern techniques, may be precisely measured with greater assurance. It is therefore not unreasonable to expect that $g_L$ may deviate from the classical value 1.0.

We remark, nevertheless, that it can indeed be rationally argued that when $\delta_L$ is set equal to zero, the difference $\xi = \delta_S - \alpha_{SL}$ between the precise $\delta_S$ and the measured $\alpha_{SL}$ is in fact the residual bound-state effects on the electron in a complex atom, which have herein been equated to $2\delta_L$. Arguments of this sort however, leave unexplained the origin of the bound-state effects which make $\xi$ in the $^2P_{1/2}$ of $Na^{23}$ as much as four times its magnitude in the $^2P_{3/2}$ of $Ga^{69}$. Moreover, we note that residual bound-state effects are typically much smaller than $\delta_L$ by several orders of magnitude. The electron g factor, which includes the Dirac value $g_D$, is given as follows [20]:

$$g = g_D + \Delta g_{QED} + \Delta g_{int} + \Delta g_{nuc} + \Delta g_{SQED} \qquad (2.2)$$

Various corrections to $g_D$ arise due to inter-electronic interaction ($\Delta g_{int}$), one-electron QED effects ($\Delta g_{QED}$), the screened QED effects ($\Delta g_{SQED}$), and nuclear effects ($\Delta g_{nuc}$). The first two quantities in the above equation are already measured in the Kusch-Foley experiment, while the next two are factored out in the analysis of their data. The other QED bound-state effects and other postulated contributions are two to five orders of magnitude smaller than the residual $\xi = \delta_S - \alpha_{SL}$, i.e, they are about $1 \times 10^{-6}$ to $1 \times 10^{-9}$ times the g-factor, as calculated for Li-like ion. Similar quantities, which have been calculated/measured for hydrogen-like $^{12}C^{5+}$, $^{16}O^{7+}$ and $^{40}Ca^{19+}$, are of the same order of magnitude [21].

**(b). Parameter-dependent Correction Factors $\delta_S$ & $\delta_L$:** More important results follow if it is observed that the two choices of $\delta_S$ and $\delta_L$ listed above are not exhaustive. It is reasonable to assume that $\delta_S$ and $\delta_L$ are inter-related and, as will be shown below, approximate results compatible (to an order of magnitude) with available measurements follow if we accept as shown [16,17], that both $g_S$ and $g_L$ are modified by a small quantity $\Delta$ such that, $g_S = 2(1+\Delta)$ & $g_L = 1-\Delta$ which implies that $\delta_S = +2\Delta$ and $\delta_L = -\Delta$. Substituting these into ($\delta_S - 2\delta_L$) measured in the first Kusch & Foley experiment gives the expression $4\Delta = 0.00229 \pm 0.00008$, from which it follows that $\Delta = 0.00057 \pm 0.00002$, and similarly, substituting them into the same quantity measured in the second experiment implies that $4\Delta = 0.00244 \pm 0.00006$ or $\Delta = 0.00061 \pm 0.00002$.

These values of $\Delta$ may be compared with the correction calculated from the non-relativistic extended Pauli equation (eqn 1,4) which, on assuming that the kinetic energy of the electron $E_k = \frac{1}{2} mv^2$ is approximately equal to $h\varepsilon$, gives a correction $\Delta = 0.4 \times 10^{-3}$ to the g-factors. It is significant, in view of the various approximations made in the calculation of $\Delta$, that its value (in the non-relativistic limit) is comparable with that obtained from Kusch & Foley data as described above.

The value of $\Delta$ ($0.61 \times 10^{-4}$) obtained above from data may be rounded up to the nearest order of magnitude to give $\Delta \simeq 0.00100$. It is interesting to note that the correction factor $\Delta$ obtained from the two experiments reported by Kusch & Foley, is nearly equal while, on the contrary, there is an unexplained discrepancy between the values of $g_S$ deduced from the experiments [12] when $\delta_S$ and $\delta_L$ are assumed to be independent and $\delta_L$ is set equal to zero. Thus we write, to an order of approximation, $\Delta \simeq 1.0 \times 10^{-3}$.

We note from the above that $\Delta$ is of the same order of magnitude as $a_e = 1.12 \times 10^{-3}$ which, to five places of decimal, is the measured anomalous part of the intrinsic g-factor. Thus, we may write

$$\Delta \simeq \left(\frac{g_S - 2}{2}\right) = a_e \qquad (2.3)$$

Hence, it is not unreasonable to assume that both $g_S$ and $g_L$ are modified as suggested in eqns.(1.6), and thus precise measurements of the $g_L$ and $g_J$ are crucial. Using (1.6), the effect of bound-state influences, if any, has been naturally included in both $\delta_S$ and $\delta_L$.

We can further the discussion by considering (1.8) where $g_S = 2(1+\Delta)$ and $g_L = 1 - \lambda\Delta$, and $\lambda$ is a real number. Putting $\delta_S = 2\Delta$ and $\delta_L = -\Delta$ into $\delta_S - 2\delta_L = \alpha_{SL}$ where $\alpha_{SL}$ = 0.00229 or $\alpha_{SL}$ = 0.00244, we have, $2\Delta(1+\lambda) = \alpha_{SL}$. Thus assuming independent $\delta_S$ and $\delta_L$, the value of $\lambda$ compatible with the Kusch-Foley data is $\lambda$ = 0.05172; this is another was of stating the observed fact that the data implies that $\delta_L \neq 0$ and thus $g_L$ cannot be exactly equal to its classical value 1.0. A more precisely measured value of $\alpha_{SL}$ will no doubt help confirm (or refute) the various points of view discussed in the preceding sections. Moreover, experiments designed to directly measure $\delta_L$ will be very significant for the quantum theories of fields and particles.

## 3. Conclusions & Discussion: Need for High Precision Measurements

By bringing together the recent high-precision measurements of the intrinsic g-factors and the classic experiment of Kusch & Foley which determined ($\delta_S - 2\delta_L$) from the ratio of the $g_J$ factors of Na and Ga atomic states (obtained from their Zeeman spectra), it has been possible to conclude that the orbital g of the electron deviates from the classical value $g_L = 1$. A probable value of the orbital g is $g_L = 1 - \delta_L$ where $\delta_L = (0.6 \pm 0.3) \times 10^{-4}$. It is however more likely that the true value of the orbital g is $g_L = 1 - a_e$, where $2a_e$ is the anomalous part of the intrinsic gyro-magnetic factor $g_S$ which has, in recent times, been measured with unparalleled precision [1,2,3].

It has been shown that an extended Dirac equation which includes the fluctuation of the rest energy term leads to a modified Pauli equation; an interpretation of the new terms in the extended Pauli equation indicates that $g_S$ and $g_L$ are corrected by finite measurable quantities. A complementary view described in the **Appendix** utilizes a simple heuristic argument in which quantum interactions, by analogy with the radiative processes in QED, are assumed to produce corrections to the electronic charge and mass, leading to the modifications of the gyro-magnetic factors $g_S$, $g_L$ and $g_J$. The corrections $\delta e$ and $\delta m$ to the charge and mass respectively, unlike the radiative corrections are, however, assumed to be finite. Further analysis indicate that corrections to both the spin and orbital g-factors are linearly related, and that an expression $\delta_S - 2\delta_L = a_{SL}$, similar to the Kusch & Foley measurement results, follows when $\delta_S$ and $\delta_L$ are assumed to be non-vanishing. Their interdependence can be understood from a physical point of view as follows; if the quantum interactions responsible for the changes in the mass and charge include spin-orbit interactions, then the orbital and spin magnetic moments may not be independent, since mutual interactions between the spins and the orbital motions of the electrons are possible. Assuming that the corrections $\delta_S$ and $\delta_L$ depend on parameter $\Delta$, then the Kusch-Foley experimental data implies that $g_S = 2(1+\Delta)$ and $g_L = 1 - \Delta$, where $\Delta$ is of the order of magnitude of the measured $(g_S - 2)/2$.

The measurement of ($\delta_S - 2\delta_L$) as described by Kusch & Foley may be subject to some limitations due to the possible perturbations of the energy levels since Ga and Na are many-electron atoms, and the effect of the system of electrons and the nuclei, may not be completely taken into account or totally eliminated. Despite these limitations however, it has been possible to consistently infer that $\delta_L$ may not vanish. One must therefore avoid assuming *a priori* that $\delta_L = 0$, and thus ascribing to the influence of systematic errors in the measurement of $\alpha_{SL}$, the observed fact that $\delta_S - \alpha_{SL} \neq 0$. Although there are well-known methods for the elimination or the reduction of systematic errors [22], the suggestion can conceivable be made that residual, systematic errors in the measurement of $\alpha_{SL}$ are what have here been called $2\delta_L$. That this reasoning is untenable may be shown as follows: Let $\alpha_{SL}$ be subject to a systematic error $\gamma$, such that $\alpha_{SL} \pm \gamma$ is the true value of ($\delta_S - 2\delta_L$). The systematic shift $\gamma$ in the measurement would have to increase in magnitude three-fold and change sign in the two experiments in order for $\delta_L$ to be equal to zero, which is contrary to the behavior of systematic errors. It therefore seems highly improbable that any systematic shift $\gamma$ in the measurement of ($\delta_S - 2\delta_L$), as determined from the Zeeman spectra displayed by two different pairs of atomic systems, would act in such a peculiar way, just so that $\delta_L$ could vanish. We may also wish to note that it could be conversely argued that the effect of systematic errors, if any, may be such that their elimination would increase $\alpha_{SL}$ to its proper value, which will make $\delta_L$ higher than the approximate value herein calculated, i.e. closer to the expected true value. Thus, only further experiments and precise, refined measurements of $(\delta_S - 2\delta_L)$ and/or $\delta_L$ can firmly establish, or otherwise repudiate, these suppositions.

One analysis of the Kusch-Foley experiment assumes the independence of $\delta_S$ and $\delta_L$ in the expressions $g_S = 2 + \delta_S$ and $g_L = 1 + \delta_L$; there is, however, no reason to assume, *a priori*, that the corrections to $g_S$ and $g_L$ are independent. As shown above, they could have a common origin and thus be related by a parameter as suggested in eqn (1.8) which, with $\lambda = 0.05172$ adequately fits the data, giving $g_L = 0.99986$ and $g_S = 2.00232$ with the latter in good agreement with the currently accepted measured & calculated values to five places of decimal. Hitherto, the expression $\delta_S - 2\delta_L = \alpha_{SL}$ had been used to determine $\delta_S$ by arbitrarily setting $\delta_L = 0$, on the basis of the "correspondence principle[3]" [23] which seems inappropriate to the situation. This procedure gave an acceptable value of $\delta_S$ sufficiently close to the value calculated at that time. Since we now have available to us, more precise values of $\delta_S$ measured by several different techniques (quantum jump spectroscopy, polarization precession method, magnetic resonance methods etc) which are reliable and in agreement with one another, we can determine the true value of $\delta_L$ from the equation $\delta_S - 2\delta_L = \alpha_{SL}$. Substituting the known, precisely measured values of $\delta_S$ into the expression and assuming that $\delta_S$ and $\delta_L$ are independent, we see that $\delta_L$ is either $(+0.2 \pm 0.4) \times 10^{-3}$ or $(-0.6 \pm 0.2) \times 10^{-3}$. If, on the other hand, $\delta_S$ and $\delta_L$ are related, then a larger value of $\delta_L$ can be obtained. For example, if $\delta_S = 2\Delta$ and $\delta_L = -\Delta$, it is found that $\Delta \sim 1.0 \times 10^{-3}$, which is of the same order of magnitude as $(g_S - 2)/2$, i.e., the anomalous part of $g_S$. However, there is, as yet, no independent precise measurement (direct or indirect) of the electron $g_L$ and/or of $\delta_L$.

Hence, the need for high precision experiments whereby $\delta_L$ may be determined, given that $\delta_S$ has been measured to extra-ordinary levels of accuracy and precision by the above-listed methods. The precise measurement of $(\delta_S - 2\delta_L)$ employing novel techniques, other that used by Kusch & Foley, will help determine $\delta_L$ and hence show whether or not $g_L = 1$ to the same high level of precision to which $g_S$ has been measured. There are currently available new methods of spectroscopy, viz Quantum beats, Doppler-free saturation spectroscopy, Level-crossing spectroscopy etc., which permit higher resolution of neighbouring levels. More accurate data and precise measurement may possibly be obtained from them.

From the preceding analysis, we see that it is not sufficiently clear or absolutely certain that $\delta_L = 0$, and that $g_L = 1$. The available evidence does not allow an incontrovertible assertion that $\delta_L = 0$, but rather supports the view that $g_L$ is possibly, by quantum effects changed from its classical value $g_L = 1$. The situation will greatly benefit from new high precision experiments, which will help clarify unambiguously the true value of $g_L$.

---

[3] From a reproduced statement of the correspondence principle [23], it may be observed that if it can be correctly argued on the strength of the principle that $g_L = 1$ and $\delta_L = 0$ for the states of Na and Ga on which measurements were carried out, then it should be equally true to regard as unquantized the orbital angular momenta and energies for those states, contrary to the physical facts.

## Appendix: Expressions Relating the Gyromagnetic-Factor Corrections

Assuming that the gyromagnetic factors have non-zero, finite values we show, using a simple heuristic argument, that $\delta_S$ and $\delta_L$ are linearly related; furthermore we derive an expression $\delta_S - 2\delta_L = a_{SL}$ which, in form, is similar to the result of the Kusch & Foley experiment. Finally, the correction $\delta_J$ to the factor $g_J$ is expressed in terms of $\delta_S$ and $\delta_L$, and the change in the magnetic dipole moment $\mu_J$, as a consequence of $\delta_J$, is discussed.

The value of the intrinsic gyro-magnetic factor $g_S = 2$ is modified by a small quantity $\delta s$ due to quantum corrections (e.g. QED radiative corrections), such that the corrected factor $g_S = 2 + \delta s$. However, the possible modification of the orbital g by $\delta_L$, and that of the total angular momentum by $\delta_J$, has received less considerations.

A heuristic argument for the modification of all the gyromagnetic factors corresponding to **S** the spin, **L** the orbital angular momentum and **J** = **L** + **S** total angular momentum, may be understood by considering the following: Let the electron magnetic moment $\mu_K$ corresponding to the angular momentum **K** be written as

$$\mu_K = \frac{eg_K}{2mc} K \tag{A1}$$

where **K** is a generic symbol for the angular momenta **L**, **S** & **J**, e/m is the measured charge-to-mass ratio, and $g_K$ is the gyro-magnetic factor corresponding to the angular momentum **K**. Let the charge and mass of the particle be respectively modified by a quantum interaction,[4] such that e → e′ = e + δe and m → m′ = m + δm, where δe and δm are, in general, finite functions of **p**, **r** and of the angular momenta, including the spin-orbit term coupling terms.

Considering the effect on the magnetic moment, of the quantum corrections of the mass and charge, we write

$$\mu'_K = \frac{e'g_K}{2m'c} K \tag{A2}$$

---

[4] The interaction is, in general, of the form $V_{p,\sigma}(r,t;\xi_j)$, where **p** is the momentum, $\sigma_i$ are the intrinsic degrees of freedom (e.g spin s), **r** & t are space & time coordinates respectively, and $\xi_j$ represents other relevant parameters. We shall be considering phenomena which are slowly varying in time, and hence have the time-dependence averaged out.

The charge-to-mass ratio $e'/m'$ can be expressed as follows:

$$\frac{e'}{m'} = \frac{e+\delta e}{m+\delta m} = \frac{e[1+\delta e/e]}{m[1+\delta m/m]} = \frac{e}{m}\left[\frac{1+\delta e/e}{1+\delta m/m}\right] \equiv \frac{e}{m} Z_K \tag{A3}$$

where, $Z_K = \{[1+\delta e/e]/[1+\delta m/m]\}$ is a function of **L**, **S** and **K**. Substituting (1.3) into (1.2) gives

$$\mu'_K = \frac{e}{2mc} Z_K g_K K \tag{A4}$$

We may therefore suppose that the effect on the $\mu'_K$, of the quantum corrections contained in $Z_K$, is to modify the gyromagnetic factor $g_K$ such that $g_K \to g'_K$. As a result, the gyromagnetic factor $g'_K$ in the presence of the quantum corrections will differ from the $g_K$ observed when the interactions considered above are absent; i.e, there are interesting consequences when $Z_K \neq 1$. Thus we may infer from eqn (A4) the expression,

$$g'_K = g_K Z_K \tag{A5}$$

Hence for **K** = **S**, $g_S \to g'_S$ and for **K** = **L**, $g_L \to g'_L$; likewise for **K** = **J**, $g_J \to g'_J$. The modification of the intrinsic g (i.e $g_s$) by non-classical interactions is well-known; experiments have shown that $Z_S = 1 + a_e$ where $a_e$ gives the anomalous part of the intrinsic magnetic moment. It is not unreasonable therefore to expect that the quantum interactions which couple spin and orbital degrees of freedom, will give rise to the mutual modifications of both the $g_L$ and $g_S$ that, in high precision experiments, may be detected without ambiguity; indeed further considerations indicate that $Z_L = 1 - a_e$. The inter–dependence of the spin and the orbital gyro-magnetic factors will be demonstrated below.

In QED, the interaction of the electron with its own radiation field, leads to a correction of the mass as follows; $m \to m' = m + ¢_1$, where the self-energy term is expressed as $¢_1 = (3m\alpha/2\pi)\ln(\Lambda/m)+...$ Similarly, vacuum polarization requires that the electronic charge e be corrected as follows; $e \to e' = e¢_3^{1/2}$, the term $¢_3$ being given as $¢_3 = 1 + (\alpha/3\pi)\ln(\Lambda^2/m)+...$ The factors $¢_1$ and $(¢_3 - 1)$ in QED are analogous to $\delta$m and $\delta$e above respectively and, as is well known, the radiative corrections in QED lead to a modification of $g_S$ by $\delta_S$ [12,24]; it should likewise be possible to calculate from QED a finite correction of the g-factor $g_L$ by $\delta_L$. The factors $¢_1$ and $¢_3$, as they occur in QED, are infinite; the theory can, nevertheless, be renormalized to accommodate the divergencies.

We shall here be concerned, however, with interactions which produce small, finite corrections (or corrections which can be made small and finite), and therefore write, from eqn (A3),

$$\frac{e'}{m'} = \frac{e[1+\delta e/e]}{m[1+\delta m/m]} = \frac{e}{m}\left[\left(1+\frac{\delta e}{e}\right)\left(1+\frac{\delta m}{m}\right)^{-1}\right]_K \quad (A6)$$

Assuming that the terms δe/e & δm/m are small compared to 1 (i.e. $\delta m/m = 1$), and linearly expanding the expression [1 + δm/m]⁻¹ in a convergent binomial series, we observe that

$$\frac{e'}{m'} = \frac{e[1+\delta e/e]}{m[1+\delta m/m]} = \frac{e}{m}\left[1+\frac{\delta e}{e}-\frac{\delta m}{m}-\frac{\delta e \delta m}{em}+....\right]_K \equiv \frac{e}{m}[1+\mathcal{Z}_K] \quad (A7)$$

from which, when substituted into eqn (A6), we can infer that

$$g'_K = g_K + g_K \mathcal{Z}_K \quad (A8)$$

since $Z_K = 1 + \mathcal{Z}_K$ from (A3) and (A7). Hence,

$$\delta_S = g'_S - g_S = g_S \mathcal{Z}_S \text{ and } \delta_L = g'_L - g_L = g_L \mathcal{Z}_L \quad (A9)$$

If $\mathcal{Z}_L = \mathcal{Z}_S$, then from (A9) $g_L \delta_S = g_S \delta_L$ which, with $g_S$ = 2 and $g_L$ = 1, implies that $\delta_S - 2\delta_L = 0$, contrary to the experimental results of Kusch and Foley. If, however, $\mathcal{Z}_S \neq \mathcal{Z}_L$ then, because $\mathcal{Z}_S / \mathcal{Z}_L = \mathcal{Z}_{SL} \neq 1$ is finite, we see that

$$\delta_S = 2\delta_L \mathcal{Z}_{SL} \quad (A10)$$

from which it follows that $\delta_S$ and $\delta_L$ are mutually interdependent. Furthermore, eqn (A10) can also be expressed as

$$\delta_S - 2\delta_L = a_{SL} \quad (A11)$$

where $a_{SL} \equiv 2\delta_L(\mathcal{Z}_{SL}-1)$, which agrees, in form, with the result of the Kusch & Foley experiment.

Let **K = J** in eqn (A2) where **J = L + S**, then the magnetic moment vector is

$$\mu'_J = \frac{e'}{2m'c} g_J J \quad (A12)$$

On the basis of the vector model [15], we assume that

$$\mu'_J = \mu'_L + \mu'_S \quad (A13)$$

then,

$$\mu'_J = \frac{e}{2mc} g_J Z_J J = \frac{e}{2mc}\left(g_L Z_L L + g_S Z_S S\right) \; ; \; \mu'_J = \frac{e}{2mc} g'_J J \qquad (A14)$$

Taking the components of the magnetic dipole moment vectors in the direction of **J** in (A14) gives,

$$g_J Z_J |J|_J = g_L Z_L |L|\cos(L,J) + g_S Z_S |S|\cos(S,J) \qquad (A15)$$

Noting that cos(L,J) and cos(S,J) are

$$\cos(L,J) = \frac{|J|^2 + |L|^2 - |S|^2}{2|J||L|} \quad \text{and} \quad \cos(S,J) = \frac{|J|^2 + |S|^2 - |L|^2}{2|J||S|}$$

equation (A15) becomes,

$$Z_J = Z_L \frac{g_L}{g_J} \frac{|J|^2 + |L|^2 - |S|^2}{2|J|^2} + Z_S \frac{g_S}{g_J} \frac{|J|^2 + |S|^2 - |L|^2}{2|J|^2} \qquad (A16)$$

where $|L|^2 = l(l+1), |S|^2 = s(s+1) \text{ and } |J|^2 = j(j+1)$. Putting

$$\alpha \equiv \frac{g_L}{g_J} \frac{|J|^2 + |L|^2 - |S|^2}{2|J|^2} \quad \text{and} \quad \beta \equiv \frac{g_S}{g_J} \frac{|J|^2 + |S|^2 - |L|^2}{2|J|^2} \qquad (A17)$$

and recalling that, in general, $Z_L = 1 - \lambda\Delta$ and $Z_S = 1 + \Delta$, where $\Delta \approx 1.12 \times 10^{-3}$, we have

$$Z_J = \alpha' - \beta'\Delta \; ; \; \alpha + \beta = \alpha', \; \lambda\alpha - \beta = \beta' \qquad (A18a)$$

or

$$Z_J = \alpha'(1 - \gamma'\Delta) \; ; \; \gamma' = (\lambda\alpha - \beta)/(\alpha + \beta) \qquad (A18b)$$

Hence, as defined in (A5),

$$g'_J = g_J Z_J = g_J \alpha'(1 - \gamma'\Delta) \qquad (A19)$$

We may, as was done for $\delta_S$ and $\delta_L$, define $\delta_J$ using $Z_J$ such that $\delta_J = g_J(Z_J - 1)$ which, upon substituting for $Z_J$ from (A18) gives $\delta_J = g_J[(\alpha' - 1) - \alpha'\gamma'\Delta]$. For a singlet state (S = 0), $\alpha' = g_J = \gamma' = 1$, hence $\delta_J = -\Delta = \delta_L$ and for L = 0, the correction $\delta_J = +2\Delta = \delta_S$, as should be expected for $\lambda = 1$. It also follows from (A9) & (A16) or from (A5) & (A15) that

$$\delta_J = \delta_L \frac{|L|}{|J|} \cos(L,J) + \delta_S \frac{|S|}{|J|} \cos(S,J) \qquad (A20)$$

Thus the contribution to $\delta_J$, from $\delta_S$ and $\delta_L$ can be reduced or enhanced by the proper choice of L, S and J. Hence the empirical consequences of $\delta_L$ may be clearly resolved in carefully controlled experiments.

It is however necessary for completeness, to make clear the difference between the $g'_J$ in (A19), and that calculated/measured in H-like [21, 26] & Li-like ions [20]. High-precision measurements of the factor $g_J$ of an electron bound in highly-charged ions have recently come into prominence, making it possible to investigate QED effects for electrons confined by strong Coulomb fields [25-30]. These measurements, being mainly concerned with the 1S½ and 2S½ states (in which L = 0), are in essence calculations/measurements of the $g_S$, i.e., the spin g-factor for bound-state electrons in strong fields. The corrected g-factors $g'_L = g_L + \delta_L$, $g'_S = g_S + \delta_S$ and $g'_J = g_J + \delta_J$ described above refer to the situation in which quantum effects are significant and the electron is weakly coupled, i.e. relatively "free"; thus, for example, the orbital g-factor for a weakly-bound atomic electron will be different from the classical value $g_L = 1$, as a consequence of the quantum-relativistic fluctuations of the rest energy. Also, $g'_J$ for such electrons (assuming LS coupling) will be different from that to be expected in the absence of quantum-relativistic effects, but will nevertheless be measurable for weakly-bound electrons after extraneous factors have been appropriately eliminated or taken into account. Hence $\delta_J$ described by (A19) will differ significantly from that arising from atomic bound-state effects due to the size & shape of the nuclear charge, nuclear-recoil corrections, electron self-energy etc. It will therefore be instructive to compare with experimental values measured for states other than the nS, the theoretical values of $g_J$ predicted by bound-state QED; electron states with non-zero orbital angular momentum are very significant for the complete determination of $g'_J$.

Finally, the substitution of (A19) into (A14) shows also that the magnetic dipole moment vector $\mu_J$ is significantly modified when $\delta_S$ and $\delta_L$ have non-zero values. The effect of the calculated correction $\delta_J$ may therefore be observed in high-precision spectroscopic or magnetic resonance measurements.